# Substrate matters: Coupled phonon modes of a spherical particle on a substrate probed with EELS

Ka Yin Lee[1], Elliot K. Beutler[2], Tifany Q. Crisolo[3], David J. Masiello[2], Maureen J. Lagos [1*]

[1] Department of Materials Science and Engineering, McMaster University, Hamilton, Ontario L8S 3N4, Canada
[2] Department of Chemistry, University of Washington, Seattle, Washington 98195, United States
[3] School of Physics, Faculty of Physical Sciences, National University of San Marcos, Lima, Lima 01, Peru

Using vibrational electron energy loss spectroscopy (vib-EELS) combined with numerical modeling, we investigate the physical mechanisms governing the phonon coupling between a spherical particle sustaining multipolar surface phonon modes and an underlying thin film. Depending upon their dielectric composition, a variety of hybrid phonon modes arise in the EEL spectrum due to the interaction between polarization charges in the particle and film. Mirror charge effects and phonon mode hybridization are the active mechanisms acting on dielectric and metallic-type films, respectively. Processes beyond dipole-dipole interactions are required to describe the sphere-film coupling.

## 1. Introduction

Local environments play a crucial role in the surface phononic response of nanoparticles [1–3]. Particularly, substrates can significantly influence the phonon polariton response of supported dielectric nanoparticles through phonon coupling [4,5]. Investigating the coupling between a nanostructure and an underlying thin film under infrared (IR) light illumination carries significant importance for advancement of fundamental knowledge in areas of IR nanophotonics, and for development of photonic applications such as radiative cooling [6], IR resonators [7,8] and sensors [9].

Theoretical approaches addressing phonon coupling in a particle-support system date back to the late 1970s [4,5]. Those studies were limited to treating the response of small particles under IR illumination in the dipole approximation, which accounts solely for dipole-dipole interactions. Under similar approximations and considering fast electrons as probes for the assessment of the IR response of nanoparticles, recent theoretical studies led to the prediction of phonon mode mixing in a nanocube supported on a dielectric slab [10,11]. Phonon hybridization between the dipole mode of a spherical nanoparticle (i.e. Fröhlich mode) and surface phonon modes of a thin film (i.e. Fuchs-Kliewer modes) [12] was also predicted. The importance of accounting for multipolar phonon modes of a spherical nanoparticle in the coupling was recently highlighted for the case of a semi-infinite support [13]. Despite this progress, a theoretical model considering a thin film as a support is still needed to assist with the interpretation of complex IR spectroscopy data obtained in transmission acquisition modes.

From an experimental perspective, the IR responses of nanostructures on a support is traditionally examined with optical-based spectroscopy techniques that probe usually the average response of an ensemble of nanostructures due to the large size of the illuminating probe. This limits in some cases the validation of predictions of phonon coupling due to close-proximity effects among nanostructures [14]. Thus, local probes with single-particle sensitivity and nanoscale resolution are imperative to assess the surface phononic response of individual nanostructures over a support and rigorously test theoretical predictions.

Vibrational electron energy loss spectroscopy (vib-EELS) in combination with scanning transmission electron microscopy (STEM) has emerged as an alternative spectroscopic technique to probe vibrational modes of isolated nanostructures with nanoscale resolution [15]. One key property of an electron beam is its ability to excite multipolar phonon modes that are typically inaccessible to far-field optical probes at vibrational energies. Thus, vib-EELS [16] offers unique opportunities to investigate new channels for energy transfer processes (i.e., coupled phonon modes involving multipolar surface phonon contributions) available in nanostructures supported on thin films.

In this work, we present a comprehensive experimental and theoretical study of the effects of a thin dielectric film support on the surface phonon response of an individual spherical dielectric nanoparticle. Specifically, we considered three sphere-film coupling scenarios:
*i*) Transparency: the film is 'transparent' ($\varepsilon_{film} \sim 1$) to the induced electric fields of the excited surface phonon modes of the sphere.
*ii*) Image charge effects: the film acts as a mirror ($\varepsilon_{film} > 1$) to the induced electric fields of the multipolar phonon modes of the sphere [17].
*iii*) Phonon-phonon coupling effects: the film acts as a metal surface ($\varepsilon_{film} < 0$) and sustains Fuchs-Kliewer phonon modes[18] that couples with the phonon modes of the sphere.
The last two cases include prominent coupling effects that are described using dielectric theory. Here, we investigate these effects using numerical finite element method simulations of the electron-sample interaction [19]. We additionally experimentally investigate the IR sphere-substrate interaction using spatially resolved vib-EELS together with numerical fitting of the resulting data with input parameters extracted from our model predictions.

## 2. Surface phonon modes of a free-space spherical particle

To begin, we investigate the phononic response of a single spherical particle suspended in vacuum (i.e., a free space sphere) using numerical simulations. The sphere is composed of silica and is 300 nm in diameter and was interrogated by a 60 keV electron beam traveling in the aloof geometry with impact parameter of 20 nm. Figure 1a shows the calculated EELS probability (black curve) associated with the excitation of surface phonon modes sustained in the sphere with each peak corresponding to an excited mode. Their phonon energies remain confined within a narrow spectral region of a few meV contained in the Reststrahlen band of the material [11].

Induced surface charge distributions for the excited phonon modes are also displayed, revealing their characteristic spatial symmetry. These multipolar modes are usually called dipole (*D*), quadrupole (*Q*), hexapole (*H*) and so on according to their number of charge pole pairs [20,21]. Additional numerical calculations revealed that the surface phonon energies can be tuned by changing the sphere size, obtaining a red shift of energies as the size increases as shown in Fig. S1.

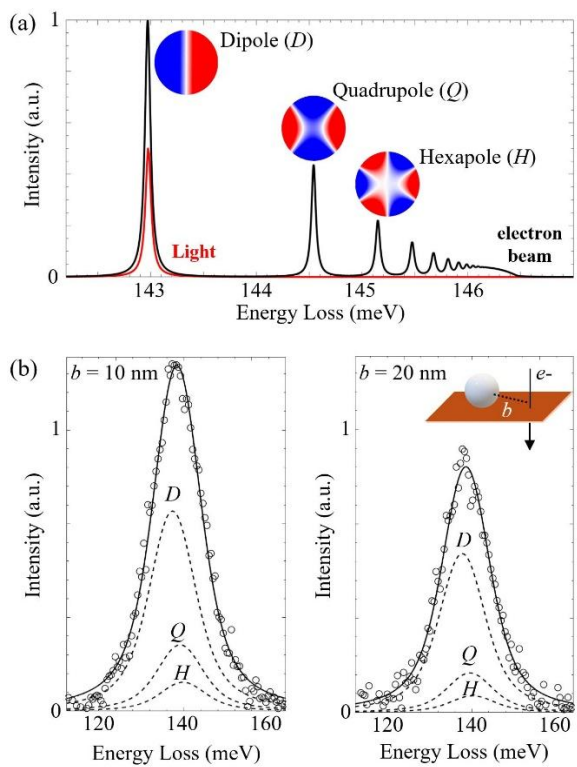

**Fig. 1.** Transparency behavior. (a) Simulated EELS probability and optical extinction cross-section spectra of a 300 nm silica free-space sphere excited by a 60 keV fast electron (black line) and IR light (red line). Induced surface charge distribution associated with the excitation of the multipolar surface phonon modes (i.e., dipole, quadrupole, etc.) are shown. Red and blue colors represent positive and negative charges, respectively. (b) Subtracted EEL spectra acquired from a 300 nm silica sphere supported on an *a*-$Si_3N_4$ thin film. Inset: Schematics of the aloof excitation of the sphere-film composite system. *b* represents the impact parameter distance. The silica sphere was probed with a 10 nm and 20 nm impact parameter. Circles: experimental data points. Dotted line: Individual contributions of multipolar modes (i.e. *D*, *Q*, *H*); higher orders are omitted for illustrative clarity. Solid line: Sum of all contributions.

Based on these calculations, it was observed that the calculated EEL spectrum is dominated by the excitation of the dipole surface phonon mode with gradual reduction in intensities of the higher order multipolar modes. We found that the high order modes tend to contribute more to the total scattering intensity at close proximity (i.e., small impact parameters)[22], while the dipole mode dominates in regimes of large impact parameters as shown in Fig. S2. Interestingly, if one probes the sphere with IR light only the dipole mode is excited, as shown by the red curve in

Figure 1a. This variation in phononic responses due to external probe conditions shows that a swift electron can access surface phonon modes with more complex charge configurations than far-field optical probes.

It should be noted that in these calculations a damping value of 0.05 meV was used for the dielectric function of silica (the actual value is 9 meV) [23]. This choice was adopted to reduce overlap of each phonon mode contribution, thus allowing a better visualization of each mode and its contribution to the overall EEL spectrum.

## 3. Transparency behavior

To study the transparency effect of the film on the phononic response of the spherical particle we selected amorphous silica ($a$-SiO$_2$) and amorphous silicon nitride ($a$-Si$_3$N$_4$) for the sphere and film materials, respectively. For this material combination the dielectric function of the nitride film is nearly 1 within the spectral regime comprising the multipolar phonon modes of the silica sphere[24] (Fig. S4a), indicating optimal conditions to evaluate the absence of coupling.

We performed vib-EELS experiments on an $a$-SiO$_2$ sphere of 300 nm in diameter supported on $a$-Si$_3$N$_4$ film of 15 nm in thickness. The sphere was probed in the aloof mode (inset of Fig. 1b) at discrete probe positions with impact parameter distances of 5, 10, and 20 nm. A broad and asymmetric EELS peak appears within each of the two Reststrahlen bands of the silica. In this study we focused on the resonances lying within the upper band around 138 meV due to its prominent appearance. Figure 1b depicts the resonance peak exhibiting an asymmetric profile due to the contribution of the high order multipolar phonon modes to the total scattering (Fig. S5). As expected, the EELS peak intensity depends on the proximity of the electron beam to the sphere, and it decreases with increase in impact parameter due to the evanescent character of the induced electric field acting on the fast electron.

To test our theoretical predictions on the free-space spherical particle, we analyzed the total contribution from the multipolar surface phonon modes and compared it to the experimental EELS resonances using a statistical fitting procedure. A detailed description of the numerical fitting implementation is presented in the appendix. The scattering contributions from each excited phonon mode convoluted with the instrumental response are indicated by the dotted curves labeled $D$, $Q$, $H$ in Figure 1b. We found an excellent fit with least square ($R^2$) values (i.e., the coefficient of determination) above 0.98 for all our data sets. The fitting parameters including the mode oscillator strength and separation between mode energies which were obtained from the analysis agree very well with values predicted by the Mie theory, indicating that the nitride film does not affect the phononic responses of the silica sphere.

To further elucidate the inactive role of the nitride film upon the silica sphere phonons, we also computed the surface charges of the supported silica sphere which is shown in Figure S7. The result reveals that the induced surface charge configurations have a strong resemblance with the ones obtained for the sphere in vacuum. Our results demonstrate that minimization of substrate effects can be achieved by carefully selecting the film permittivity similar to the vacuum permittivity at the nanoparticle's resonant energy range.

## 4. Fundamentals of surface phonon coupling

In this section, we present the fundamentals of the phonon coupling between a spherical particle and a thin film of finite thickness. The effects of a film on the surface phonon modes of a sphere depends on the film's dielectric properties. The physics of the coupling is based on the Coulomb interaction between the charge configurations sustained in the sphere and the film. Figure 2 illustrates the key aspects behind the phonon coupling for cases in which the film behaves as dielectric mirror ($\varepsilon_{film} > 1$) and a metallic-type surface ($\varepsilon_{film} < 0$).

For the dielectric mirror case ($\varepsilon_{film} > 1$), the two planar boundaries of the film act together to create multiple image charges which interact with the free space modes of the sphere by either *self-interaction* or *cross interaction*. This description is inspired by the approach presented in references [25–30]. Figure 2a illustrates the involved interactions between the phonon modes and their corresponding mirror charge images. As a result of this interaction, a new set of coupled mixed phonon modes are created which are distinct from the multipolar phonon modes of the free space sphere. The *cross-interaction* involves phonons of different mode orders (e.g., dipole - mirror quadrupole). Each sphere mode configuration can be classified as polarization charges within the plane (e.g., $D_i$, $Q_i$, where $i = x, y$ represents orthogonal in-plane directions) and out of the plane (e.g., $D_z$, $Q_z$). The coupling between modes and their mirror images is more favorable for charge polarizations that tend to align than the ones that are orthogonal [31].

The energy states of the resultant mixed modes for the dielectric mirror case can be visualized in the energy level diagram shown in Fig. 2b. For coupling between in-plane modes, $D_i^+$ and $D_i^-$ are the in and out-of-phase modes resulting from the *cross-interaction* between a free space dipole and a quadruple mode image, respectively. $Q_i^+$ and $Q_i^-$ are the in and out-of-phase mixed modes between a free space quadrupole and a hexapole image, respectively. $D_i'$ is the result of in-phase *self-interaction* of $D_i$ with its own image. There is no out-of-phase mode for the *self-interaction* because their spatial charges cancel. Similarly, for coupling between out-of-plane modes, $D_z^+$ and $D_z^-$ are the in-phase and out-of-phase modes resulting from the cross-interaction between a free space dipole and a quadrupole mode image, respectively. $D_z'$ is the result of in-phase *self-interaction* of $D_z$ with its own image.

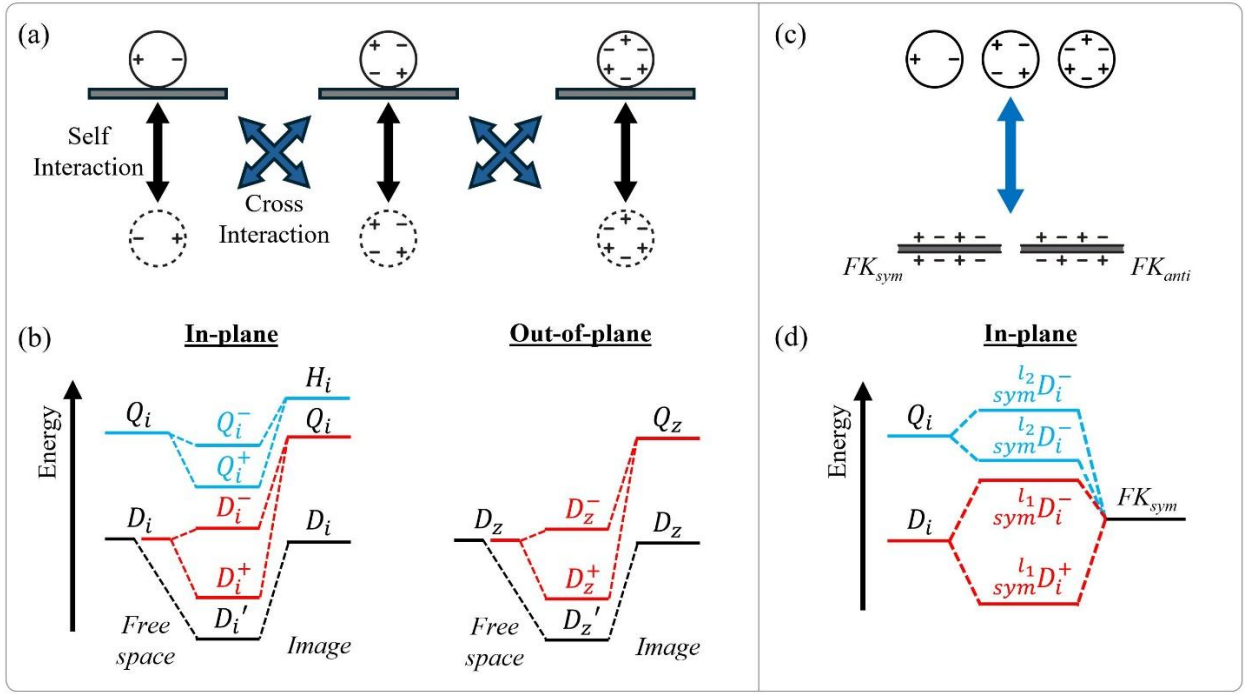

**Fig. 2.** Fundamentals of phonon mode mixing and phonon-phonon coupling. (a) Charge schematics illustrating the multipolar surface phonon mode mixing pathways due to the image charge effect. (b) Energy level diagram of the mixed modes as compared to the free space and image modes of the sphere. The subscript $i = x, y$ represents the two orthogonal in-plane directions, and $z$ represents the out-of-plane directions. The self interactions of the higher order modes are omitted for clarity. (c) Charge schematics illustrating the phonon coupling between phonon modes sustained in the sphere and the film (symmetric/anti-symmetric Fuchs-Kliewer modes). (d) Energy level diagram illustrating the energy level of the coupled modes. See text for explanation.

For the case of a metallic-type thin film ($\varepsilon_{film} < 0$), the film can sustain symmetric ($FK_{sym}$) and anti-symmetric ($FK_{anti}$) Fuchs-Kliewer modes that disperse within the Reststrahlen band of the material [18]. Figure 2c shows the schematics of charge configurations for the FK modes of the film and the multipolar phonon modes (e.g., $D$, $Q$, $H$) sustained by the sphere. The symmetric and anti-symmetric FK modes have charge polarizations aligning parallel and perpendicular to the film surfaces, respectively [11]. The coupling between phonon modes in the sphere and film is more efficient when their polarizations tend to align, and the energy detuning parameter approaches zero. For instance, the in-plane charge polarization components ($D_i$, $Q_i$) of the sphere couple more efficiently to the $FK_{sym}$ mode than the $FK_{anti}$ due to the orientation of their in-plane polarizations.

As a result of their interaction, coupled phonon modes are created with different energy states as illustrated in the energy level diagram shown in Figure 2d. $^{l_1}_{sym}D_i^+$ and $^{l_1}_{sym}D_i^-$ are the in-phase and out-of-phase modes of the coupling involving the dipole mode and the symmetric FK, respectively. While $^{l_2}_{sym}D_i^+$ and $^{l_2}_{sym}D_i^-$ are the in-phase and out-of-phase modes resulting from the coupling between the quadrupole and symmetric FK, respectively. The energy separation of dipole-FK coupled modes is larger than that of the quadruple-FK coupled modes, indicating a stronger coupling of the former. For modes with out-of-plane polarization, the energy of the multipolar modes and the anti-symmetric FK modes are detuned from each other, so their coupling is weaker than the in-plane coupling mode, which is revealed in the reduced energy shift relative to their free space counterparts (Fig. S8).

## 5. Image charge effects: phonon mode mixing

To investigate the image charge effect of a dielectric film on the surface phonon modes of a spherical particle, we selected amorphous silicon nitride ($a$-$Si_3N_4$) and magnesium oxide (MgO) for the materials making up the film and sphere, respectively. For this combination, the real part of the dielectric function of the nitride film is 7.6 [24] in the spectral band comprising the multipolar phonon modes of the MgO sphere [23] (Figure S4b). Strong mirror effects are obtained given the large values of the dielectric function of the nitride film (i.e. $\varepsilon_{film} \gg 1$). A small absorption component around 2 plays a role in the damping of modes and other combinations of materials with smaller permittivity were also explored, which yields similar outcomes (results are not shown).

To begin the analysis of the coupling effect, we consider a scenario in which the coupling gradually enhances as the distance between the sphere and film reduces. Thus, we performed EELS simulations considering a MgO sphere of 300 nm in diameter located at a certain distance above the nitride film. The sphere is probed by a 60 keV electron in the aloof mode with a 20 nm impact parameter. The silicon nitride film thickness is 15 nm. The results are shown in Fig. S9. For large separation distances (> 100 nm), the sphere's dielectric response is similar to that in vacuum due to the weak electrostatic sphere-film interaction. As the sphere approaches the film, the *self-interaction* and *cross interaction* (Fig. 2a) gradually increases leading to the generation of mixed surface phonon modes. The energy separation between the modes is driven by the strength of the Coulomb interaction modulated by the permittivity of the film. Larger permittivity values lead to stronger mode separation.

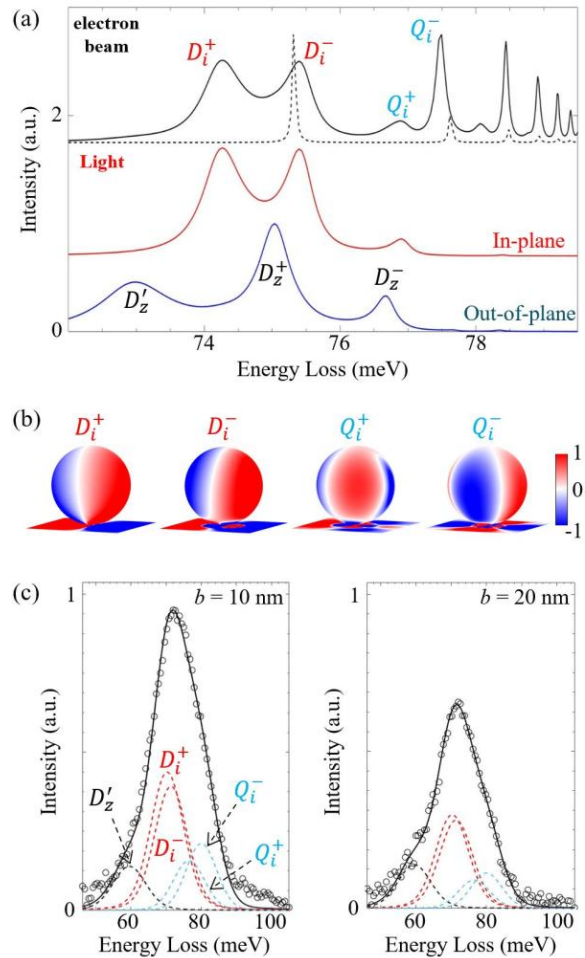

**Fig. 3.** Coupled phonon modes due to mirror charge effects. (a) Simulated EELS and optical extinction cross-section spectra of a MgO sphere of 300 nm in diameter at 5 nm above an $a$-Si$_3$N$_4$ thin film under electron beam excitation (black curve), and light excitation with in-plane (red curve) and out-of-plane (blue curve) polarization. Simulated EELS of a free-space MgO sphere (dotted black curve) is also shown. (b) Simulated surface charge configurations of the most prominent mixed surface phonon modes. Charge values are normalized to the maximum of each map. (c) Subtracted EEL spectra acquired for a single MgO sphere supported on an $a$-Si$_3$N$_4$ film with electrons travelling with 10 nm and 20 nm impact parameter. Circle: experimental points. Dotted line: Individual mixed phonon contributions. Solid line: Sum of all contributions. Higher order contributions are omitted for clarity.

The black curve in Figure 3a shows the simulated EELS probability of the MgO sphere at 5 nm above the $a$-Si$_3$N$_4$ film. For comparison we also displayed the simulated EELS probability of an isolated free-space MgO sphere of the same size (black dotted curve). A close inspection reveals that the EELS peaks for the coupled system are quite distinct from the multipolar phonon modes of the free-space sphere. We identified the mixed dipole-type modes ($D_i^-$, $D_i^+$) and the mixed quadrupole-type modes ($Q_i^+$, $Q_i^-$) resulting from *cross-interaction* processes, which were discussed in the previous section. The simulated scattering intensity of $Q_i^-$, $D_i^-$ and $D_i^+$ dominate the spectrum followed by the nearly undisturbed higher order multipolar modes (above the hexapole mode).

To identify the orientation of the charge polarizations of the mixed coupled modes, we performed simulations using plane-wave light illumination under parallel (*in-plane*) and perpendicular (*out-of-plane*) polarization conditions (Fig. 3a). When the composite sphere-film system is illuminated using *in-plane* polarization, $D_i^-$, $D_i^+$ and $Q_i^+$ are solely excited with the dipole-type components dominating the whole spectrum. For *out-of-plane* illumination, the $D_z'$, $D_z^+$ and $D_z^-$ modes are excited. As mentioned above, the $D_z'$ mode is the result of the *self-interaction* of the out-of-plane dipoles appearing at the low energy side. The $D_z^+$ and $D_z^-$ have polarization along the direction perpendicular to the film, thus favoring this coupling. The comparison between the simulated spectra indicates that the swift electron couples more strongly to mixed phonon modes with charge polarizations lying parallel to the film surface than those with perpendicular orientation. This coupling is likely favored due to the flattened spatial distribution of the electric field of the relativistic electron [32].

Figure 3b shows simulated surface charge maps on both the sphere and film (top surface) for the most prominent mixed phonon modes appearing in the simulated EEL spectrum shown in Fig. 3a. $D_i^+$ shows a dipolar configuration dressing the sphere. Near the closest contact point with the film (i.e. proximal side) the surface charge increases drastically. The film surface mirrors the charge configuration on the sphere. For $D_i^-$, the dipole pattern over the sphere remains nearly the same as $D_i^+$, but a reversal of its charge configuration is built on the proximal side. A dipole charge pattern is visualized on the film surface mirroring the sphere configuration. On the other hand, the surface charge pattern of $Q_i^+$ exhibits a quadrupolar nature which is highly localized near the contact point accompanied with a weaker in-plane quadrupole over the sphere surface. The surface charges of $Q_i^-$ are in opposite phase when compared to that of $Q_i^+$. For both modes, the charge on the film adopts a mirroring pattern of the charge configuration on the sphere.

To evaluate the validity of these predictions we performed vibrational EELS experiments. We probed a MgO sphere of 300 nm in diameter supported on $a$-Si$_3$N$_4$ film of 15 nm in thickness with impact parameter 5, 10, and 20 nm. Figure 3c shows the subtracted EEL spectra acquired at 10 nm and 20 nm impact parameter. A strong asymmetric resonance appears at about 76 meV. A shoulder-like spectral feature appears on the lower energy side at around 60 meV, becoming more visible at large impact parameters. We analyzed the total contribution from the mixed phonon modes ($D_z'$, $D_i^+$, $D_i^-$, $Q_i^+$, $Q_i^-$, …) and compared it to the EELS peaks through a numerical fitting. The contribution for modes above $Q_i^-$ were modeled as one single peak as their energy separations are well below 0.1 meV. Most of the scattering signal around 76 meV is accounted for the in-plane $D_i$ and $Q_i$ mixed modes, while the 60 meV contribution is attributed to the out-of-plane $D_z'$ mode. Notice that the individual resonances cannot be resolved due to the limited spectral resolution. Also, the broad contribution of each mode is due to the instrumental response (~ 10 meV). We also found that the scattering associated with each mode decays gradually as the impact parameter increases (Fig. S3).

An excellent fit with R$^2$ values around 0.99 for all the fitted set of data was obtained, revealing strong consistency of the derived fitting results. We also noticed that if one considers only the low-energy mode contributions (e.g. $D_i^+$, $D_i^-$) then it is not possible to fit satisfactorily the experimental data. This demonstrates that mixed phonon modes of high order nature are quite active, and theoretical approaches beyond dipole-dipole (self) interactions are indeed needed to describe the phonon coupling process of the composite sphere-film system.

## 6. Phonon-phonon coupling

To investigate the phonon coupling between the Fuchs-Kliewer (FK) phonon modes of the thin film and multipolar phonon modes of the spherical particle, we selected amorphous silica ($a$-SiO$_2$) as the material making up both the sphere and film. The dielectric function of the silica is negative within the spectral band of the phonon modes revealing their metal-like response [23].

We performed EELS simulations considering a silica sphere of 300 nm in diameter located at certain distance above a silica film (Fig. S10) with the film thickness being 15 nm. The composite sphere-film system is probed by a 60 keV electron traveling with a 20 nm impact parameter. As expected, the coupling between the FK modes and multipolar modes gradually increases as the sphere gets closer to the film leading to the hybridization of phonon modes. Figure 4a (black solid line) displays the simulated EELS probability for the sphere at 5 nm above the film. In contrast to the general red-shift behavior imposed by the mirror-like film on the sphere, we found blue shifts of the EELS resonances associated with the excitations of dipole-FK ($^{l_1}_{sym}D_i^-$, $^{l_1}_{anti}D_z^-$) and quadrupole-FK modes ($^{l_2}_{sym}D_i^-$, $^{l_2}_{anti}D_z^-$). A magnified view of the $^{l_2}_{sym}D_i^-$, $^{l_2}_{anti}D_z^-$ is depicted in the figure inset. Notice that these two resonances are very close to each other, with $^{l_2}_{anti}D_z^-$ appearing as a shoulder in the higher energy tail of $^{l_2}_{sym}D_i^-$. We observe that the EELS probability involving quadrupolar modes is about 40% of that of the dipolar modes, which highlights the importance of accounting for contributions beyond dipole-dipole interactions.

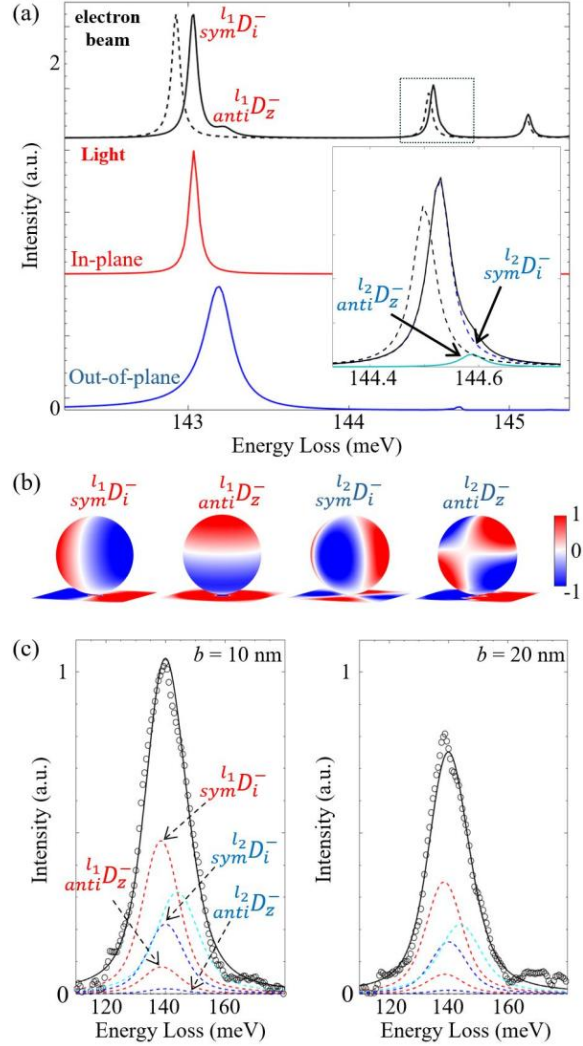

**Fig. 4.** Coupled phonon modes due to surface phonon interactions. (a) Simulated EELS and optical extinction cross-section spectra of a silica sphere of 300 nm in diameter located at 5 nm above a silica film under electron beam excitation (black curve) and light excitation with in-plane polarization (red line) and out-of-plane polarization (blue line). Simulated EEL spectrum of a free-space silica sphere is shown (dotted black curve). Inset: Amplified view of the rectangular region marked on (a) highlighting coupled quadrupole-FK modes. (b) Simulated surface charge maps of the coupled phonon modes. Charge values are normalized to the maximum of each map. (c) Subtracted EEL spectra acquired for a single silica sphere on silica thin film using electrons traveling with 10 and 20 nm impact parameter. Solid circle: Experimental data points. Dotted line: Individual mode contributions. Cyan line indicates the total contribution of higher order modes above hexapole mode. Solid line: Sum of all contributions.

To identify the orientation of the charge polarizations of the coupled phonon modes, we computed the optical extinction cross section of the composite system illuminated under in-plane (red curve) and out-of-plane (blue curve) polarizations. We found that $^{l_1}_{sym}D_i^-$ is efficiently excited under in-plane polarization illumination and $^{l_1}_{anti}D_z^-$ modes under out-of-plane illumination. Higher energy modes are less likely to be excited using optical illumination. Modes with charge polarizations lying predominantly parallel to the substrate plane (e.g. $_{sym}^{l_1}D_i^-$, $_{sym}^{l_2}D_i^-$) are more efficiently excited when compared to the ones with perpendicular polarizations.

Figure 4b shows simulated surface charge maps of the coupled modes. Notice that $_{sym}^{l_1}D_i^-$ mode exhibits an in-plane dipolar pattern on the sphere surface accompanied with the typical symmetric FK charge pattern on the film, but with reversed charge polarity with respect to the sphere. $_{anti}^{l_1}D_z^-$ mode displays a dipolar pattern oriented along the perpendicular direction accompanied with the anti-symmetric FK mode on the film mirroring the sphere charge configuration. The polarizability of these modes was also worked out analytically in the quasistatic limit[12] and our simulation results are in mutual agreement. Furthermore, $_{sym}^{l_2}D_i^-$ mode exhibits an in-plane quadrupolar pattern bound to a mirror pattern developed on the film top surface. The $_{anti}^{l_2}D_z^-$ mode also displays a four-fold symmetry pattern with an in-plane dipole configuration on the lower hemisphere of the sphere. A mirror image pattern is built on the film top surface.

To validate our predictions, we probed a silica sphere of 300 nm in diameter supported on a silica film of 15 nm in thickness using a 60 keV electron beam traveling with impact parameters ranging between 5 and 20 nm. Figure 4c shows the subtracted EEL spectra acquired at 10 and 20 nm impact parameters. An asymmetric resonance appears at about 140 meV. The scattering intensity of the peak decays as the impact parameter increases.

To assess the contributions of the coupled hybrid modes to the experimental scattering, we numerically fit their contributions to the set of EELS data. The contribution for modes above $_{anti}^{l_2}D_z^-$ were modeled as one single component as their energy separations are well below 0.1 meV. Our fitting results revealed excellent $R^2$ values well above 0.98. To achieve optimal fittings conditions, components including quadrupole interactions were needed. Otherwise, a proper fitting is not possible when only accounting for dipole-dipole interactions such as the $_{sym}^{l_1}D_i^-$, $_{anti}^{l_1}D_z^-$ modes.

Furthermore, by examining the scattering EELS intensities of the hybrid coupled modes as a function of impact parameter, we notice that the in-plane hybrid scattering signal decay faster as compared to the out-of-plane hybrids as the probe moves away from the sphere (Fig. S3).

We also compared the spectral behavior of the silica sphere coupled to the thin film with the nearly-free silica sphere studied in the transparency scenario (Fig. 1b) to examine the blue shift response of the EELS peaks. We found a 3 meV spectral shift of the broad peak towards higher energies for the coupled sphere with respect to the uncoupled sphere. As evidenced by our results, this apparent shift is a result from the redistribution of the scattering signal towards higher energies due to the excitation of higher order coupled phonon modes.

## 7. Summary

We investigated the physics of surface phonon coupling between a spherical particle and a thin film under different coupling scenarios. Through a combination of experimental vibrational EEL measurements and numerical simulations, our work revealed a variety of different coupled phonon channels on the composite sphere-film system in which the fast electron can deposit energy during an inelastic scattering event. For a dielectric film ($\varepsilon_{film}$ >1), phonon-mode mixing takes place assisted by mirror charge effects, while for metallic-type films ($\varepsilon_{film}$ < 0) phonon-phonon coupling occurs by favorable tuning conditions of the modes sustained in the sphere and film. Coupled phonon modes appear in the spectra as red or blue spectral shifts when compared to the behavior displayed by a free-space sphere. To minimize or eliminate the phonon coupling in the composite system (i.e., transparency behavior), the dielectric function of the film can be selected near to unity within the spectral range of the surface phonons sustained in the sphere.

Spatially resolved vibrational EELS measurements conducted in a single supported sphere validated our theoretical predictions at different probing conditions. We also demonstrated that a complete treatment of phonon coupling in the sphere-film composite system requires theoretical modeling accounting for interactions beyond dipole-dipole approximations. Similar considerations are applied to studies of supports with different morphologies.

Furthermore, our findings reveal that despite the reduced thickness of the thin film (below 15 nm) it can have a great impact on the dynamics of coupled surface phonons, altering significantly the overall infrared response of a composite sphere-film system, which in turn would have potential impact on the performance of IR photonic applications that rely on nanostructures supported on thin films.

Our work presents a comprehensive study of phonon coupling scenarios, extending our understanding of the physics of coupled phonon modes of isolated nanostructures supported on thin films and revealing the significance of the substrate contribution. Also, these results bring important physical insights into the interpretation of complex vibrational EELS data.

**Acknowledgments**

We acknowledge the Natural Sciences and Engineering Research Council of Canada (NSERC) under a Discovery Grant. M. J. L also acknowledges the Canada Research Chair Program for providing funding support. Work at the University of Washington was supported by the U.S. Department of Energy (DOE), Office of Science, Office of Basic Energy Sciences (BES), Materials Sciences and Engineering Division under Award DE-SC0022921 (D.J.M.). We thank V. Amarasinghe for providing silica film supports and P. E. Batson for providing access to the Rutgers Nion UltraSTEM to acquire part of the vib-EELS data. We also thank the Canadian Centre for Electron Microscopy (CCEM) for providing access to electron microscopy facilities.

**Appendix A: Sample preparation**

1. $a$-SiO$_2$ Sphere on an $a$-Si$_3$N$_4$ thin film

The $a$-SiO$_2$ spheres were purchased from Alpha Nanotech. They were prepared by the sol-gel process based on the Stöber method and came in the form of $a$-SiO$_2$ nanoparticles suspended in a solution. The dispersed $a$-SiO$_2$ nanoparticle solution (300 nm in diameter) was drop casted on an $a$-Si$_3$N$_4$ thin membrane and dried in air. The 10 nm thick silicon nitride membranes were purchased from Norcada Inc. The part number is NT050Z. The membrane was synthesized via lower-pressure chemical vapor deposition (LPCVD).

2. MgO Sphere on an $a$-Si$_3$N$_4$ thin film

The MgO sphere (~ 300 nm in diameter) was prepared by first burning a magnesium wire in air, and then collecting the generated MgO smoke using a 10 nm thick silicon nitride membrane purchased from Norcada Inc. Spheres of different diameters were generated.

3. $a$-SiO$_2$ Sphere on an $a$-SiO$_2$ thin Film

We grew $a$-SiO$_2$ films on sodium chloride (NaCl) crystals using a Plasma-Therm 790 PECVD system at ~10 nm/min and 250°C. The oxide thickness was calibrated by depositing on a piece of silicon under the same conditions and measuring the thickness using optical ellipsometry. The film thickness is about 15 nm. After the deposition, the process chamber was cooled down to room temperature and the silica film was floated off the substrate onto the surface of de-ionized water and captured on a TEM grid. The dispersed $a$-SiO$_2$ spheres solution (300 nm in diameter) purchased from Alpha Nanotech was drop casted on the $a$-SiO$_2$ films and dried in air.

**Appendix B: STEM/EELS experiments**

Vibrational EELS spectra were acquired in a monochromated Nion UltraSTEM electron microscope with a ~ 1.5 Å probe at 60 kV. Energy resolution is about 10 meV. Each spectrum acquisition was acquired using a 1-2 pA electron probe with a convergence semi-angle of 30 mrad and collection semi-angle of 20 mrad. A dispersion of 0.9 meV per channel was used.

**Appendix C: EELS data acquisition and processing**

Raw spectra will consist of a stack of 2D EELS images. The stack of images will be first aligned according to the position of maximum electron count pixel in each frame and then integrated to become a single 2D image. After that, an integration window is selected along the non-dispersive direction on the single 2D image by picking 60 – 75% of the maximum electron count pixel over the entire 2D image. The resultant EELS spectra will be an 1D array of counts against energy loss. Alignment of the energy axis to the zero-loss position is done by picking 5 – 10 electron count values around the maximum electron count position in the 1D array and fitted them to a Gaussian function, so that the maximum intensity position can be determined and set as zero loss position in the energy axis.

Point EELS spectra were acquired in pairs, one with the electron probe parked next to sphere (sample spectrum), and another one parked in a space with only the thin film present (reference spectrum), away from any sphere. The reference signal was subtracted from the sample spectrum resulting in a spectrum containing scattering from the coupled sphere-film system alone, as imposed by the principle of superposition [12].

**Appendix D: Simulated EELS of a free-space sphere**

Aloof beam electron energy loss probability $\Gamma^{loss}(\omega)$ for an isotropic sphere of radius $R$ was analytically calculated using equation 37 in reference [33] derived from Mie inelastic electron scattering theory [34]. The scattering probability expression is reproduced below

$$\Gamma^{loss}_{sphere}(\omega) = \frac{1}{c\omega} \sum_{l=1}^{\infty} \sum_{m=-l}^{l} K_m^2\left(\frac{\omega r}{v\gamma}\right) \times [C^M_{lm}|t^M_{lm}|^2 + C^E_{lm}|t^E_{lm}|^2] \quad (1)$$

which is implemented in a Matlab package - metal nanoparticle boundary element method (MNPBEM20) toolbox [19], where $\gamma = (1-\beta^2)^{-\frac{1}{2}}$, $\beta = \frac{v}{c}$, r = b + R, $b$ is the impact parameter, $v$ is the speed of the electron, $c$ is the speed of light in vacuum, $K_m$ is the modified Bessel function of order $m$, $l$ is the order of the surface mode, $t^M_{lm}$ and $t^E_{lm}$ are magnetic and electric coefficients derived from Mie scattering theory respectively,

$$t^M_l = \frac{-j_l(\rho_0)\rho_1 j'_l(\rho_1) + \rho_0 j'_l(\rho_0) j_l(\rho_1)}{h^+_l(\rho_0)\rho_1 j'_l(\rho_1) - \rho_0 [h^+_l(\rho_0)]' j_l(\rho_1)} \quad (2)$$

$$t^E_l = \frac{-j_l(\rho_0)[\rho_1 j_l(\rho_1)]' + \varepsilon(\omega)[\rho_0 j_l(\rho_0)]' j_l(\rho_1)}{h^+_l(\rho_0)[\rho_1 j_l(\rho_1)]' - \varepsilon(\omega)[\rho_0 h^+_l(\rho_0)]' j_l(\rho_1)} \quad (3)$$

$j_l$ are spherical Bessel functions, $h^+_l$ are spherical Hankel functions, $\rho_0 = \frac{\omega R}{c}$, $\rho_1 = \frac{\omega R \sqrt{\varepsilon(\omega)}}{c}$, $\varepsilon(\omega)$ is the dielectric function of the sphere material,

$$C^M_{lm} = \frac{1}{l(l+1)}\left|\frac{2mv}{c} A^+_{l,m}\right|^2,$$

$$C^E_{lm} = \frac{1}{l(l+1)}\left|\frac{1}{\gamma} B_{lm}\right|^2,$$

$$B_{lm} = A^+_{l,m+1}\sqrt{(l+m+1)(l-m)} - A^+_{l,m-1}\sqrt{(l-m+1)(l+m)},$$

$$A^+_{l,m} = \frac{(-1)^m (2l+1)!!}{\beta^{l+1}} \sqrt{\frac{(2l+1)(l-m)!}{4\pi(l+m)!}} \times$$

$$\sum_{j=m}^{l} \left\{ \frac{i^{l-j} \int_{-1}^{1} d\mu (1-\mu^2)^{\frac{j}{2}} \mu^{l-j} P_l^m(\mu)}{(2\gamma)^j (l-j)! \left(\frac{j-m}{2}\right)! \left(\frac{j+m}{2}\right)!} \right\},$$

, where the summation for $A^+_{l,m}$ runs over even $(j+m)$ integers, $P_l^m$ are the Legendre functions.

**Appendix E: Simulated EELS of a composite sphere-film system**

Electron energy loss probabilities for the composite system were calculated using equation 9.5 following [32]

$$\Gamma^{loss}_{composite}(\vec{R}_0, \omega) = \frac{e}{\pi \hbar \omega} \times$$

$$\int_{-\infty}^{\infty} \text{Re}\left\{ e^{-i\omega t} \vec{v} \cdot \vec{E}_{ind}\left[\vec{r}_e(t), \omega\right] \right\} dt$$

under dielectric theory within MNPBEM20, where $\vec{r}_e(t) = \vec{R}_0 + \hat{z}vt$ represents the electron trajectory, $\vec{v}$ is the velocity vector of the fast electron. The induced electric field ($\vec{E}_{ind}$) along the beam trajectory is numerically computed by solving the boundary value problem with the full Maxwell equations in the vector and scalar potential formalism under the Lorentz gauge. To check for convergence, simulation results are compared for various number of the nanosphere's surface discretization from about 2000 to 6000 according to [35,36]. The effect of the thin film substrate is modelled based on the reflected Green's function formulated by [37] and implemented into the MATLAB package by [38]. Surface charge maps are extracted from the boundary elements when the continuity conditions of the electric and magnetic fields are applied, which is a built-in method in MNPBEM20. Extinction cross sections including the thin film substrate are computed using MNPBEM20 according to [39,40].

**Appendix F: Fitting analysis**

We used a function representing the sum of Voigt functions to perform a least square fit to our experimental data. The functions are in the form of

$$P^{EELS}_{fit}(\omega) = A \sum_{l=1}^{N} c_l \pi \gamma_l f_l^{Voigt}(\omega_l; \sigma_l, \gamma_l, E_{l,peak})$$

, where $f^{Voigt}(\omega; \sigma, \gamma_l, E_{peak})$ is the Voigt function defined by a convolution of a Lorentzian function with a Gaussian function as follows:

$$f^{Voigt}(\omega; \sigma, \gamma_l, E_{peak}) = \int_{-\infty}^{\infty} L[(\omega - E_{peak}) - x'; \gamma_l] G(x'; \sigma) dx'$$

, where $E_{peak}$ represents the energy peak positions, $L$ is the Lorentzian functions and $G$ is the Gaussian function, $A$ is a scaling parameter, $c_l$ is the fitting parameter that represents the relative contribution of the mode(s), $l$ is the order of the modes, *N* is the total no. of Voigt functions used in the fit. In this scheme, the Lorentzian and gaussian functions represent the EELS resonance and instrumental broadening, respectively. Additional information of this fitting methodology is presented in section 6 of the supplementary material.

The initial input parameters, e.g. the peak positions and separations, intensity ratios, are directly extracted from the MNPBEM20 simulated result. The minimization between the fitting function and the experimental data is performed by an interior point algorithm [41,42] offered as a package by Matlab.

**Data availability**

Data is available upon request.